\documentclass[twocolumn,showpacs]{revtex4}

\usepackage{dcolumn}
\usepackage{amsmath,amssymb}

\ifx\pdfoutput\undefined
  \usepackage[dvips]{graphicx}
\else
  \pdfoutput=1
  \usepackage[pdftex]{graphicx}
\fi
\begin{document}
%\preprint{}
\bibliographystyle{revtex}

\title{The effect of thermal annealing on the properties of \\
Al-AlO$_x$-Al single electron tunneling transistors}

\author{H. Scherer}
\email{hansjoerg.scherer@ptb.de}
\author{Th. Weimann}
\author{A. B. Zorin}
\author{J. Niemeyer}
\affiliation{Physikalisch-Technische Bundesanstalt, Bundesallee
100, 38116 Braunschweig, Germany}

\date{\today}

\begin{abstract}
The effect of thermal annealing on the properties of Al-AlO$_x$-Al
single electron tunneling transistors is reported. After treatment
of the devices by annealing processes in forming gas atmosphere at
different temperatures and for different times, distinct and
reproducible changes of their resistance and capacitance values
were found. According to the temperature regime, we observed
different behaviors as regards the resistance changes, namely the
tendency to decrease the resistance by annealing at $T = 200$~
$^\circ$C, but to increase the resistance by annealing at $T =
400$~$^\circ$C. We attribute this behavior to changes in the
aluminum oxide barriers of the tunnel junctions. The good
reproducibility of these effects with respect to the changes
observed allows the proper annealing treatment to be used for
post-process tuning of tunnel junction parameters. Also, the
influence of the annealing treatment on the noise properties of
the transistors at low frequency was investigated. In no case did
the noise figures in the $1/f$-regime show significant changes.
\end{abstract}

%Single-electron tunneling, Coulomb blockade, tunneling, metal-insulator-metal structures, single electron devices

\maketitle

\section{Introduction}

Within the last decade, single electron tunneling (SET) devices
have been intensively investigated and have demonstrated their
great potential for unique applications \cite{SET92,Likharev99}.
For instance, SET transistors may be used as ultrasensitive
electrometers of a minimal charge resolution of less than $10^{-5}
e/\sqrt{\mbox{Hz}}$ at 10 Hz \cite{Krup00}. These rather simple
SET devices consist of two ultrasmall tunnel junctions,
characterized by parameters such as their tunnel resistance $R$
and their junction capacitance $C$, connected in series by a
metallic island being capacitively coupled to a gate electrode
with capacitance $C_{\rm G}$.

In this paper, we present a study showing that a suitable forming
gas annealing (FGA) process can be used to tune the junction
properties of conventional Al-AlO$_x$-Al SET transistors in a
controllable way. Recently, several groups have extensively
studied the effects of thermal annealing on $\mu$m-scaled
Nb-Al/AlO$_x$-Nb Josephson junctions with regard to the junction
properties \cite{gates84, shiota92, lehnert92, oliva94}. However,
adequate attention has so far not been paid to similar studies on
tunnel devices comprising nm-scaled Al-AlO$_x$-Al junctions.

We had two motives in processing metallic SET transistor samples
by FGA: Firstly, we aimed at investigating the influence of a
thermal annealing procedure on the tunnel junctions with respect
to $R$ and $C$. Here, the forming gas simply served as a
protective atmosphere to prevent the unwanted excessive oxidation
of the tunnel junctions. Secondly, forming gas annealing performed
at the end of CMOS process lines in the semiconductor industry, is
a traditional method for of healing process-induced defects by the
passivation of electrically active traps in the Si/SiO$_2$
substrate material \cite{wolf86, Koizuka98, sanden99, sanden00}.
Also, it is well known that the typical $1/f$-like excess charge
noise, measured in the low-frequency regime of conventional SET
transistors, is caused by the motion of background charges in the
dielectric surroundings of the transistor islands (see, for
example, \cite{Zor96, Krup00} and references therein). Hence, it
was our special interest to study the effect of the FGA treatment
on the noise behavior of our samples that were also prepared on
Si/SiO$_2$ substrates.

\section{Experiment}

We prepared standard SET transistors by the conventional shadow
evaporation technique on an Si substrate wafer with a thermally
oxidized SiO$_2$ surface layer (300~nm thick). The three-layer
mask was made of PMMA/Ge/copolymer. After patterning of the PMMA
layer using e-beam exposure and a developing process, the pattern
was transferred to the Ge layer by an etching process in a CF$_4$
plasma, followed by oxygen plasma etching of the copolymer layer
creating the undercut for the following shadow evaporation. Next,
two layers of aluminum were deposited by e-gun evaporation under
different angles, and {\em in situ} oxidation of the bottom Al
layer (200~Pa O$_2$, 10~min) between the evaporation steps formed
the Al-AlO$_x$-Al tunnel junctions. Each of the sample chips
carried eight transistors whose junction sizes were intentionally
varied on each chip while the island dimensions were kept constant
(Fig.~1).
\begin{figure}
  \centering
  \includegraphics[scale=0.3,clip=true]{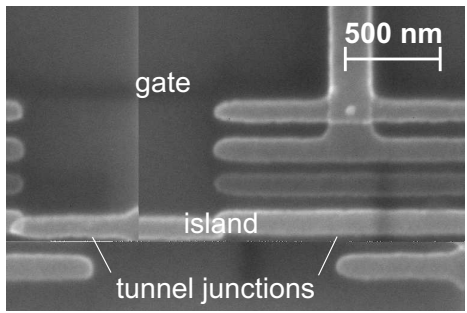}
  \caption{SEM picture of a typical SET transistor after FGA at 400 $^\circ$C
  for 45 min. No changes are seen in comparison with the picture before annealing.
  The areas of the tunnel junctions are about $80 \times
  80$ nm$^2$, the total initial island capacitance $C_{\Sigma,\rm 0} =  500$ aF,
  and the total initial resistance $R_{\Sigma,\rm 0} = 380$ k$\Omega$.
 }
\end{figure}

The sample chips were annealed in an electrical furnace at a
temperature of 200~$^\circ$C or 400~$^\circ$C in a forming gas
atmosphere of 10~\% H$_2$ and 90~\% N$_2$ (by volume) at a total
pressure $\approx 800$~hPa. Before and between successive
annealing cycles, the furnace chamber was flushed with forming gas
and evacuated several times to minimize the amount of residual
oxygen in the system that entered the system when it was opened to
the air. Annealing time dependent resistance changes of the
samples were traced by multimeter measurements under normal
conditions, performed between successive annealing cycles. The low
temperature characterization of the samples was performed in a
top-loading dilution refrigerator at $T = 25$~mK and $B = 1$~T
before and directly after FGA treatment \cite{Footnote1}. All
signal lines were equipped with Thermocoax cables about 1 m in
length, serving as microwave frequency filters \cite{Zor95}, and a
low-noise electronic setup, consisting of a symmetrical current or
voltage bias and low-noise preamplifiers, was used for the
measurements. Noise spectra were recorded using an HP 89410A
spectrum analyzer. The total resistance value $R_{\Sigma} \approx
2 R$ of a sample was given by the asymptotic differential
resistance. The external impedance of the aluminum lines connected
to the island was rather low, so the total capacitance
$C_{\Sigma}$ was derived by extrapolating the linear part of the
``offset plot'', i.e. $U_{\rm off} \equiv U - I dU/dI$ versus $U$,
to zero voltage \cite{Wahl95, Wahl98}.

\section{Results and discussion}

All samples investigated survived the annealing treatment, as it
was checked by multimeter measurements directly after FGA. In
particular, even after annealing at 400 $^\circ$C for 45 min, the
SET transistors retained their typical gate modulation properties
without any indication of qualitative degradation. Some samples
were also inspected using an electron microscope before and after
the FGA (Fig.~1), but no visible changes in their shape or surface
morphology due to the annealing process were detected. However,
distinct quantitative changes in the main electrical parameters of
all samples occurred that are described in the following.

\subsection{Junction parameters after FGA at 200 $^\circ$C}

A series of 27 transistor samples with junctions of different size
($R_{\Sigma, \rm 0} \approx 100$~k$\Omega$ - 800~k$\Omega$,
$C_{\Sigma,\rm 0} \approx 1.1$~fF - 0.25~fF) was exposed to FGA at
a temperature of 200~$^\circ$C for 15 min to 120 min. 16
transistor samples were characterized at 25 mK before and after
the FGA. Unexpectedly, we observed a decrease of the sample
resistance $R_{\Sigma}$ during annealing.
\begin{figure}
  \centering
  \includegraphics[scale=0.4,clip=true]{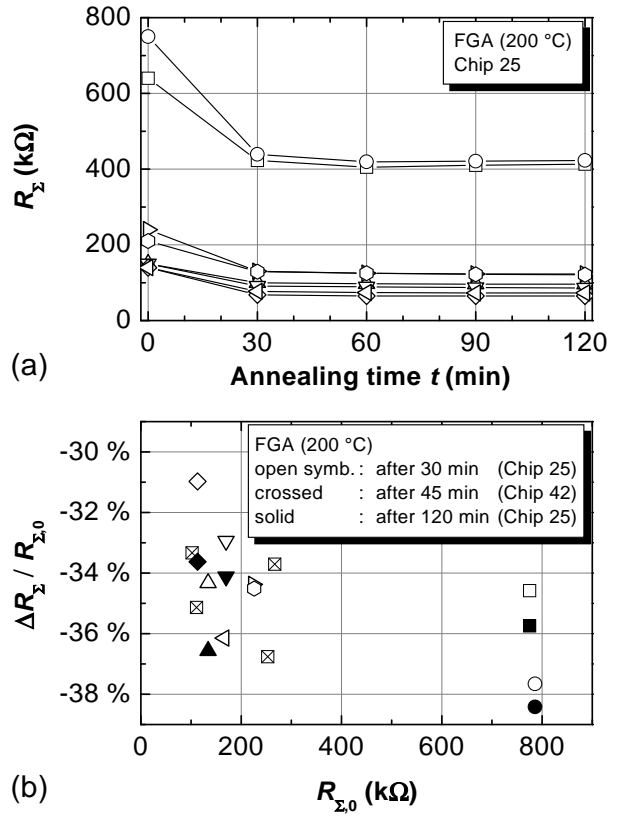}
  \caption{(a) Effect of decreasing sample resistances after FGA at
  200 $^\circ$C for different annealing times with the typical
  behavior of resistance saturation after 30 min of annealing. (b)
  Relative change $\Delta R_{\Sigma}/R_{\Sigma, \rm 0} =
  (R_{\Sigma}-R_{\Sigma, \rm 0})/R_{\Sigma, \rm 0}$ for several
  samples versus their initial resistance values $R_{\Sigma, \rm 0}$.
  Resistance values are given in units of k$\Omega$, and the
  shapes of the symbols correspond to individual transistor samples.
 }
\end{figure}
The typical behavior is demonstrated in Fig.~2, showing the
relative resistance change $\Delta R_{\Sigma}/R_{\Sigma, \rm 0}
\equiv (R_{\Sigma}-R_{\Sigma, \rm 0})/R_{\Sigma, \rm 0}$ versus
$R_{\Sigma, \rm 0}$, and the saturation of the resistance change
after about 30 min. For all samples, we found a quite similar
change of the saturated $\Delta R_{\Sigma}/R_{\Sigma, \rm 0}$ of
about 30~\% - 40~\%. This behavior was reproduced for different
sample chips and annealing runs.

We verified that the junctions retained pure tunnel contact
behavior after the annealing treatments by making sure that all
samples showed a distinct Coulomb blockade regime with a zero
current state. Thus, we conclude that the resistance decrease is
associated with a reduction of the effective barrier thickness. On
the basis of a simple plate capacitor junction geometry where any
change in the junction resistance is due to changes in the tunnel
barrier composition, and/or thickness, we calculated the tunnel
resistance, assuming a rectangular energy barrier of height $\phi$
and thickness $s$ \cite{Simmons63}. With $\phi = 1.9$~eV - 2.6~eV
for the thermally oxidized aluminum oxide barrier
\cite{Gundlach71, McBride74} and with an initial thickness $s_{\rm
0} = 0.7$~nm - 0.8~nm (to fit the $R_{\Sigma, \rm 0}$ values), we
estimated a barrier thickness change $\Delta s/s_{\rm 0} = -4$~\%,
corresponding to the saturated resistance change. In the plate
capacitor model, this relative thickness change should be
reflected by a similar relative capacitance increase of the
junctions. Although the corresponding offset plots seemingly
indicate no change in $C_{\Sigma}$ and, respectively, in $C$
(Fig.~3), we cannot rule out such a small change within the
uncertainty due to data scattering.
\begin{figure}
  \centering
  \includegraphics[scale=0.3,clip=true]{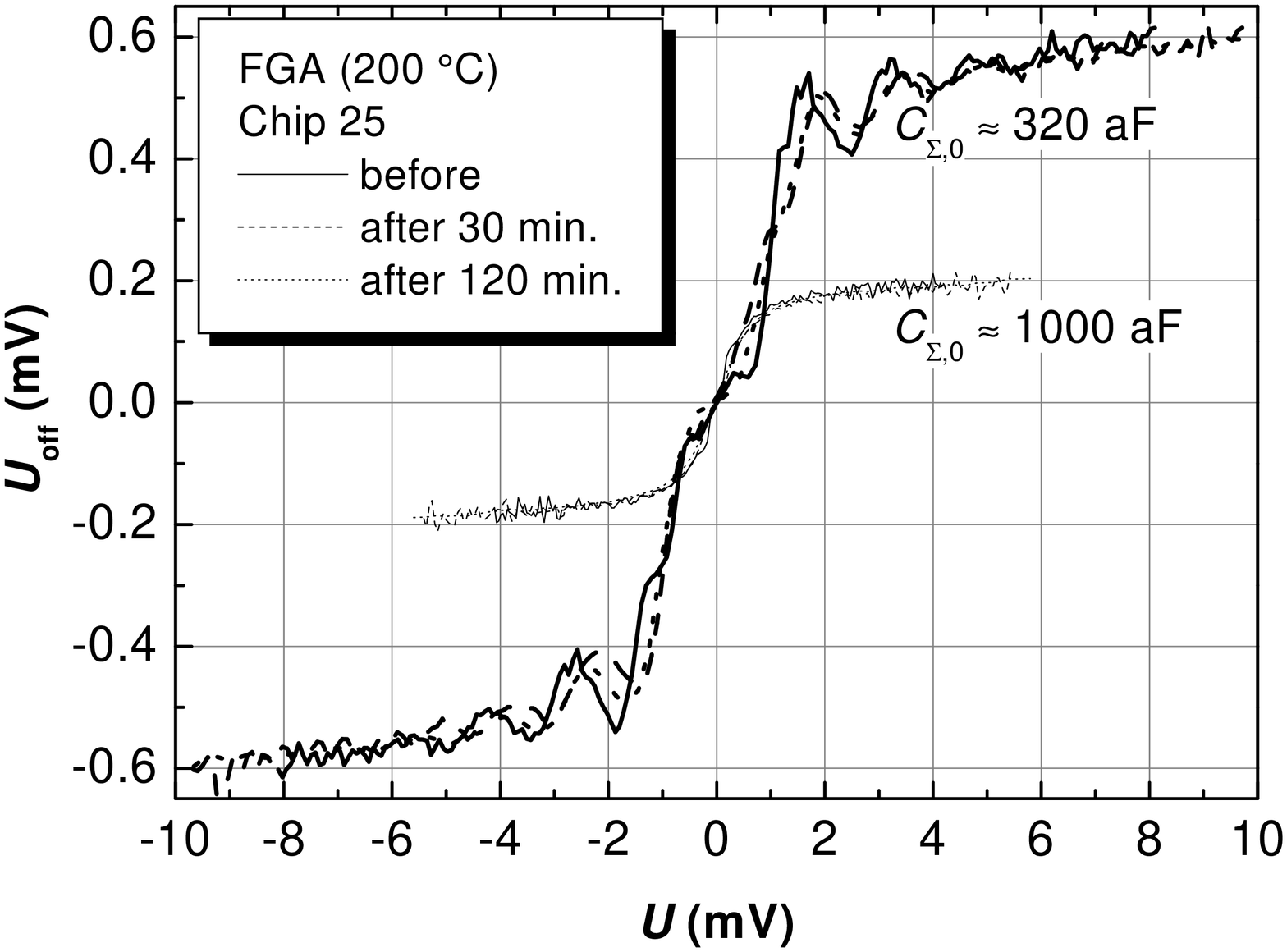}
  \caption{Offset plot for two transistors with different junction
sizes, illustrating that FGA at 200~$^\circ$C seemingly caused no
changes to the voltage offset $U_{\rm off}$ of either device ($T =
25$ mK, $B = 1$ T).}
\end{figure}

In principle, finding a junction thickness decrease due to thermal
annealing seems contrary to intuition and is in conflict with
earlier results for thermal annealing effects on Nb-Al/AlO$_x$-Nb
Josephson junctions \cite{gates84, shiota92, lehnert92, oliva94}.
Here, it was found without exception that annealing treatments in
the range from $T = 0$ $^\circ$C to 350 $^\circ$C  resulted in an
increase in the normal state junction resistance, indicating an
increase in thickness of the AlO$_x$ barriers. As in our case, the
AlO$_x$ barriers were thermally oxidized but the materials and the
deposition methods for the junction fabrication were different
(sputtering in the case of the niobium junctions, but e-gun
evaporation of the aluminum devices), which might lead to the
differences in the annealing behaviors.

Although we have at present no direct method to investigate the
microscopic changes in the barrier, we may speculate about the
scenario during annealing: Firstly, in the presence of water,
aluminum is known to form many hydroxides from Al$_2$O$_3
\cdot$H$_2$O ({\em diaspore}, AlO(OH)) to Al$_2$O$_3 \cdot$3H$_2$O
({\em gibbsite}, AlO(OH)$_3$) and substances of intermediate
composition \cite{Mellor60}. Although the initial pressure in our
evaporation chamber was $\approx 10^{-5}$~Pa, a small residual
water vapor content, and, thus, a deviation of the barrier
composition from pure Al$_2$O$_3$ cannot be ruled out. The
different hydroxide phases vary in crystalline structure, and
dehydration temperatures of 145 $^\circ$C and more are reported
\cite{Mellor60}. Even at moderate temperatures of 200~$^\circ$C,
the barrier morphology may therefore be affected by
re-crystalization processes or changes in composition, possibly
decreasing the effective barrier thickness. Although our data to
not provide a clear indication about an associated increase in
junction capacitance, even the case of $C \approx C_{\rm 0}$ and
$R < R_{\rm 0}$ is possible if the mean barrier thickness is
unchanged ($\bar{s} \approx s_{\rm 0}$), but small sharp spikes
form on the electrode surface (for instance due to
re-crystalization processes). Such spikes, ``pricking" into the
barrier, could focus the current distribution on small areas with
reduced barrier thickness and, due to the exponential dependence
of the tunnel resistance on this parameter, reduce the total
resistance.

Of course, in addition to the modifications of the barrier
morphology, dehydration processes or changes in composition would
affect the barrier height $\phi$ in an unpredictable way. Also,
the relaxation of stress inside the layers might influence the
interface morphology. We thus may state that our phenomenological
observations are distinct and reproducible, but a detailed
microscopic explanation is speculative at present.

\subsection{Junction parameters after FGA at 400 $^\circ$C}

\begin{figure}
  \centering
  \includegraphics[scale=0.4,clip=true]{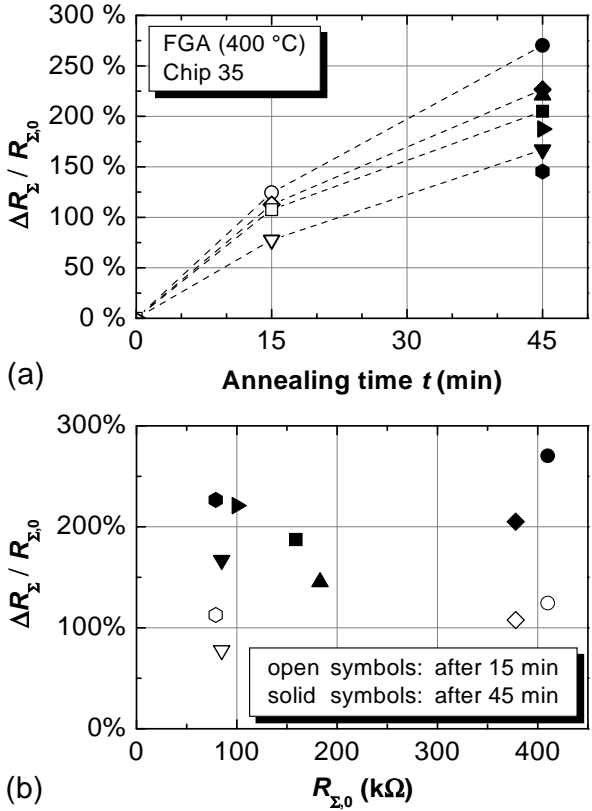}
  \caption{Dependence of the relative sample resistance increase
$\Delta R_{\Sigma}/R_{\Sigma, \rm 0}$ on the annealing time (a)
for FGA at 400~$^\circ$C and versus the initial resistance
$R_{\Sigma, \rm 0}$ (b). Resistance values are given in units of
k$\Omega$, and the shapes of the symbols correspond to individual
transistor samples.}
\end{figure}
14 transistors with the initial sample parameters $R_{\Sigma, \rm
0} \approx 80$~k$\Omega$ - 410 k$\Omega$ and $C_{\Sigma,\rm 0}
\approx$ 1.4~fF - 0.4~fF were exposed to FGA at a temperature of
400~$^\circ$C. For all samples we found that the resistance values
had approximately doubled after 15~min, while after 45~min they
were about three times $R_{\Sigma, \rm 0}$ (Fig.~4(a)). Again,
this effect was independent of the initial resistance $R_{\Sigma,
\rm 0}$ (Fig.~4(b)).
\begin{figure}
  \centering
  \includegraphics[scale=0.3,clip=true]{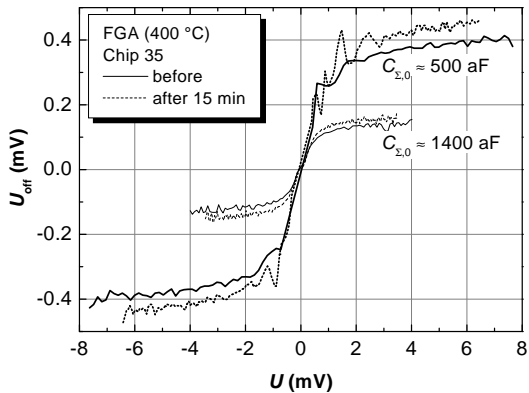}
  \caption{Offset plot for two different transistors before and
after 15 min of FGA at 400~$^\circ$C ($T = 25$~mK, $B = 1$~T).
Clearly, the voltage offset $U_{\rm off}$ has increased after
annealing.}
\end{figure}

Also, the offset plots of these samples showed a significant
increase in $U_{\rm off}$ after the annealing (Fig.~5), reflecting
a decrease of $C_{\Sigma}$. Assuming that $C_{\Sigma} = 2 C +
C_{\rm G} + C_{\rm S}$, where $C_{\rm S} \approx$ 20~aF - 40~aF is
the stray capacitance of the 1~$\mu$m long transistor islands
\cite{Wahl98, Lu98}, we estimated the mean capacitance $C$ of the
junctions for each transistor \cite{footnote3}. Fig.~6 shows the
relative decrease of the tunnel capacitance $\Delta C/C_{\rm 0}
\equiv (C-C_{\rm 0})/C_{\rm 0}$ versus $C_{\rm 0}$.
\begin{figure}
 \centering
  \includegraphics[scale=0.3,clip=true]{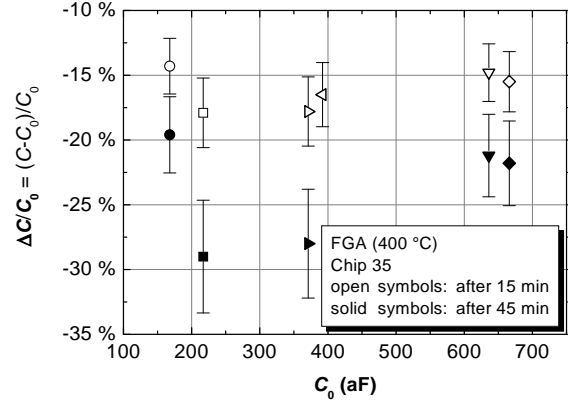}
  \caption{Reduction of the mean tunnel capacitance $C$ for
different transistors after 15~min and 45~min of FGA at
400~$^\circ$C, represented as the relative change $\Delta C/C_{\rm
0}$, versus the initial values $C_{\rm 0}$.}
\end{figure}
After 15~min of annealing, the junction capacitances had decreased
about 15~\% - 20~\%, and after~45 min about 20~\% - 30~\% from
their initial values.

These results clearly indicate that the effective oxide barrier
thickness increases as a result of the FGA treatment. In
accordance with papers \cite{gates84, shiota92, lehnert92,
oliva94}, we attribute the annealing induced oxide barrier growth
by the reaction with unbound oxygen, coming either from
interstitial lattice sites in the proximity of the junctions and
diffusing - thermally activated - towards the barrier, or directly
from dissociation of aluminum hydroxides in the barrier.

As described in the previous section, we estimated the relative
increase in thickness corresponding to the change in resistance
after 45~min, and found $\Delta s/s_{\rm 0} \approx 9$~\%. Since
this is significantly less than the value derived from the
capacitance analysis, we conclude that the simple plate capacitor
model for the tunnel junctions breaks down and that a non-uniform
change of the barrier thickness has taken place along the junction
cross section. In principle, such non-uniform changes may easily
explain our findings due to the different functional dependence of
tunnel resistance $R$ and junction capacitance $C$ on the distance
between the tunneling electrodes. Nevertheless, our study does not
enable us to make more precise statements about the microscopic
changes in the barriers.

\subsection{Low-frequency noise behavior}

Si wafers with a thermally oxidized SiO$_2$ surface layer are the
most commonly used substrate material, not only for the
preparation of SET devices but also in semiconductor industry.
There, the CMOS devices are typically processed by FGA at 400
$^\circ$C for about 30 min, since this treatment is especially
suitable to heal defects in the Si/SiO$_2$ interface (see, for
instance, \cite{wolf86, Koizuka98, sanden99, sanden00} and
reference therein). It is known that so-called interface charge
traps are located at the Si/SiO$_2$ interface, able to exchange
carriers with the Si. FGA leads to a hydrogen passivation of these
defects so that the interface density of states is typically
reduced by one order of magnitude \cite{wolf86, risch00} and the
device performance is improved generally. Recently, it has been
shown that an FGA process similar to that we carried out reduced
the power of the low frequency $1/f$ noise in bipolar junction
transistors by a factor of five, which indicates that the
interface traps were effectively passivated by hydrogen
\cite{sanden00}.

\begin{figure}
  \centering
  \includegraphics[scale=0.4,clip=true]{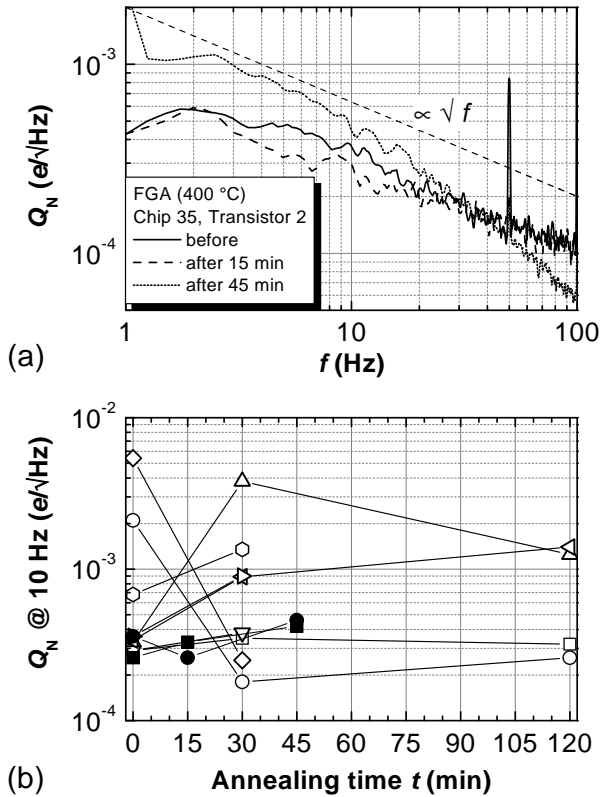}
  \caption{(a) Typical $1/f$-like charge noise spectra of a
transistor before and after different durations of FGA at
400~$^\circ$C ($T = 25$~mK, $B = 1$~T, $I = 50$~pA). (b) Annealing
time dependence of the charge noise figures $Q_{\rm N}$ at $f =
10$~Hz for several samples after FGA at 200~$^\circ$C (open
symbols) and 400~$^\circ$C (solid symbols).}
\end{figure}
Using our SET transistors prepared on such Si/SiO$_2$ substrate as
ultrasensitive electrometers, we tried to figure out the effects
of FGA on the noise figures in the low frequency regime, i.e. in
the range $f = 1$~Hz to 100 Hz. Before and after annealing at
200~$^\circ$C and 400~$^\circ$C, we found the typical $1/f$-like
noise power spectra, indicating the activities of independently
switching noise sources in the dielectric surroundings of the
transistor islands. By checking that the noise signal was
proportional to the gain in the working point of each transistor,
we could make sure that the noise sources were acting on the input
of the electrometer, which justified the conversion of the
measured noise signals into effective charge noise $Q_{\rm N}(f)$
(see Fig.~7(a)). The annealing time dependence of the charge noise
figures $Q_{\rm N}$ at $f = 10$~Hz for a several transistors
(after FGA at 200~$^\circ$C and 400~$^\circ$C) is demonstrated in
Fig.~7(b). We generally found that the noise figures varied after
thermal cycling between the different characterization runs, but
the changes showed no monotonic trend. This is the usual behavior
of SET devices when thermal cycling alters the potential
``landscape" in which the charge traps are frozen in. Moreover,
all values of $Q_{\rm N}$(10 Hz) were found to be in the range
between some $10^{-4} e/\sqrt{\mbox{Hz}}$ and some $10^{-3}
e/\sqrt{\mbox{Hz}}$, which is the typical noise level for
conventional SET transistors (see, e.g., \cite{Zor96, Krup00}, and
references therein).

Our experiments indicate that FGA processing has no significant
effect on the level of the $1/f$-like charge noise in SET
transistors, which is in contrast to the recent experiments on the
CMOS bipolar junction transistors \cite{sanden00}. The reason
probably is that hydrogen passivation mainly affects noise sources
in the Si/SiO$_2$ interface, which is 300~nm below the
noise-sensing devices in our case, whereas the charge traps
dominating the total noise in our SET devices are located closer
to the transistor island, for instance on the substrate surface.

\section{Conclusion}

We investigated the effect of thermal annealing in a forming gas
atmosphere on the properties of Al-AlO$_x$-Al SET transistors and
found distinct and reproducible changes of the junction
parameters. After annealing at $T = 200$~$^\circ$C, the sample
resistance $R_{\rm \Sigma}$ decreased by about 30~\% - 40~\%,
while the total capacitance $C_{\rm \Sigma}$ remained nearly
unchanged. After annealing at $T = 400$~$^\circ$C, we found an
increase in $R_{\rm \Sigma}$, associated with a reduction of
$C_{\rm \Sigma}$. In both cases, the effects are attributed to
changes in morphology, composition and/or effective thickness in
the oxide barriers of the junctions, which are believed to be
independent of the composition of the protective atmosphere.
Moreover, the changes obtained appeared to be stable, i.e. within
several weeks no degradation back to the initial sample values
$R_{\rm \Sigma, 0}$ and $C_{\rm \Sigma, 0}$ was observed.

In addition, the influence of forming gas annealing on the noise
properties of the transistors at low frequency was investigated.
Here, we found no significant reduction of the noise level in the
$1/f$-regime due to potential hydrogen passivation of charge
traps.

The good reproducibility of the observed effects suggests that a
proper thermal annealing treatment may be used for safe
post-process tuning of the junction parameters in SET devices.
Firstly, annealing in a protective atmosphere at $T \approx
400$~$^\circ$C may be used to increase the tunnel resistance $R$
in an easily controllable way, for instance to enhance SET effects
in devices with insufficient $R_{\rm 0} \lesssim R_{\rm K} = h/e^2
\approx 25.8$~k$\Omega$. Secondly, annealing at $T \approx
200$~$^\circ$C, causing a tunnel resistance drop of about 30~\% -
40~\%, might be useful to enhance the Josephson coupling energy
$E_J \propto R^{-1}$ in the junctions of SET devices that should
also offer superconducting properties (as, for instance, the Bloch
transistor \cite{ZorinBloch96}) but are on the borderline of their
performance due to the initial sample parameters.

\bibliography{scherer01}

\begin{thebibliography}{10}
\bibitem{Footnote1}
\bibinfo{note}{The samples were cooled down in the dilution refrigerator to $T
  = 1.5$ K within the next two hours after the last annealing treatment. They
  were kept at this temperature for about 12 hours before their low-temperature
  characterization at 25 mK.}
\bibitem{footnote3}
\bibinfo{note}{$C_{\rm G}$ was measured before and after annealing (it remained
  unchanged), and any eventual change of the small parameter $C_{\rm S}$ should
  not substantially affect the result.}
\end{thebibliography}

\end{document}